\documentclass{article}
\usepackage{spconf,amsmath,graphicx}
\usepackage{multirow}
\usepackage{tikz}
\usepackage{enumitem}
\usepackage{booktabs}
\usepackage{csquotes}
\usepackage{caption,subcaption}
\usepackage{calc}

\usepackage{spconf,amsmath,graphicx}
\usepackage{amsmath,graphicx}
\usepackage{amssymb}
\usepackage{multirow}
\usepackage{xspace}
\usepackage{xcolor}
\usepackage{textgreek}

\newcommand{\yun}[1]{\iffalse{\textcolor{cyan}{#1}}\fi}
\usepackage{algpseudocode}
\usepackage{algorithm}
\usepackage{amsfonts}

\usepackage{bm}
\usepackage{array}
\usepackage{cite}
\usepackage{multirow}
\usepackage{units}
\def\mv#1{\boldsymbol{{#1}}}
\def\mv#1{\boldsymbol{{#1}}}
\def\sxs#1#2{\ensuremath{{\text{#1}}\hspace{-1pt}{\times}\hspace{-1pt}{\text{#2}}}}%

\title{AGADIR: Towards Array-Geometry Agnostic Directional Speech Recognition}
\name{
 Ju Lin$^*$, Niko Moritz$^*$, Yiteng Huang$^*$, Ruiming Xie, Ming Sun, Christian Fuegen, Frank Seide$^{\dagger}$ 
\thanks{*Equal contribution.  {$\dagger$}Corresponding author.}}
\address{Meta, USA }

\begin{document}
\ninept

\maketitle

\begin{abstract}

Wearable devices like smart glasses are approaching the compute capability to seamlessly generate real-time closed captions for live conversations. We build on our recently introduced directional Automatic Speech Recognition (ASR) for smart glasses that have microphone arrays, which fuses multi-channel ASR with serialized output training, for wearer/conversation-partner disambiguation as well as suppression of cross-talk speech from non-target directions and noise.

When ASR work is part of a broader system-development process, one may be faced with changes to microphone geometries as system development progresses. 

This paper aims to make multi-channel ASR insensitive to limited variations of microphone-array geometry. We show that a model trained on multiple similar geometries is largely agnostic and generalizes well to new geometries, as long as they are not too different. Furthermore, training the model this way improves accuracy for seen geometries by 15 to 28\% relative. Lastly, we refine the beamforming by a novel Non-Linearly Constrained Minimum Variance criterion.

\end{abstract}


\section{Introduction}
\label{sec:intro}
Automatically transcribing a conversation partner at a distance of several feet is an important emerging ASR scenario. Consider a wearable device that automatically generates captions for deaf or hearing-impaired users. Background noise, reverberation, overlapping speech, and interfering speakers make this challenging. To remedy, one can capture the speech with a microphone array—like we humans do with binaural hearing. Microphone-array methods traditionally aim to improve the SNR of target speech---but one can do better by multi-channel Automatic Speech Recognition (ASR).

This paper extends our recently proposed directional speech-recognition system for real-time closed captions of conversations on smart glasses. That model receives multiple beamformed signals simultaneously, allowing the ASR model itself, in an end-to-end fashion, to disambiguate who is speaking between the wearer, the conversation partner, and unrelated bystanders, while also being more noise-robust than ASR on single-channel beamformed signals~\cite{lin2023directional}.

This paper aims to make the multi-channel model less sensitive to minute details of the specific microphone-array geometry, striving for Array-Geometry Agnostic Directional Speech Recognition, or AGADIR. Why? On smart glasses, the mic array competes with other components in terms of space and other considerations. During system development, consecutive prototypes tend to undergo alterations of microphone placement. A multi-channel ASR model that is agnostic to limited geometry changes could be shared across a sequence of prototypes, e.g.~for user studies, saving time and energy consumption. It would allow predicting system accuracy for new configurations without new test data. Our experiments on both simulated and real test data show that a model that is simply trained on multiple similar geometries is indeed agnostic to limited geometry variations and even leads to better WER (although it finds its limits for larger geometry changes).

Related work on geometry agnosticity includes~\cite{taherian2022one}, which proposes a causal geometry-agnostic multi-channel speech enhancement system that leverages speaker embeddings and spatial features serving as the front-end for speech recognition. An array geometry-agnostic speech separation neural network model named VarArray, was proposed in~\cite{yoshioka2022vararray}, which could be seamlessly integrated into diverse array configurations for streaming multi-talker ASR in~\cite{kanda2023vararray}.

MIMO-speech~\cite{chang2019mimo} is a multichannel end-to-end neural network that defines source-specific time-frequency masks as latent variables in the network, which in turn are used to transcribe the individual sources. This was improved by incorporating an explicit localization sub-network. Recent studies \cite{wang2018multi,shao2022multi} in ASR and speaker separation have investigated direct incorporation of spatial features instead of using explicit sub-modules jointly trained with the ASR module. For example, \cite{chen2018multi} proposed to estimate a target-speaker mask with multi-aspect features to extract the target speaker from a speech mixture. The extracted speech is then fed to ASR. Recently neural beamforming was also explored for multi-channel ASR \cite{sainath2016factored,he2020spatial}.

\section{Directional ASR System Architecture}
\begin{figure*}[!ht]
    \centering
    \includegraphics[width=0.95\textwidth]{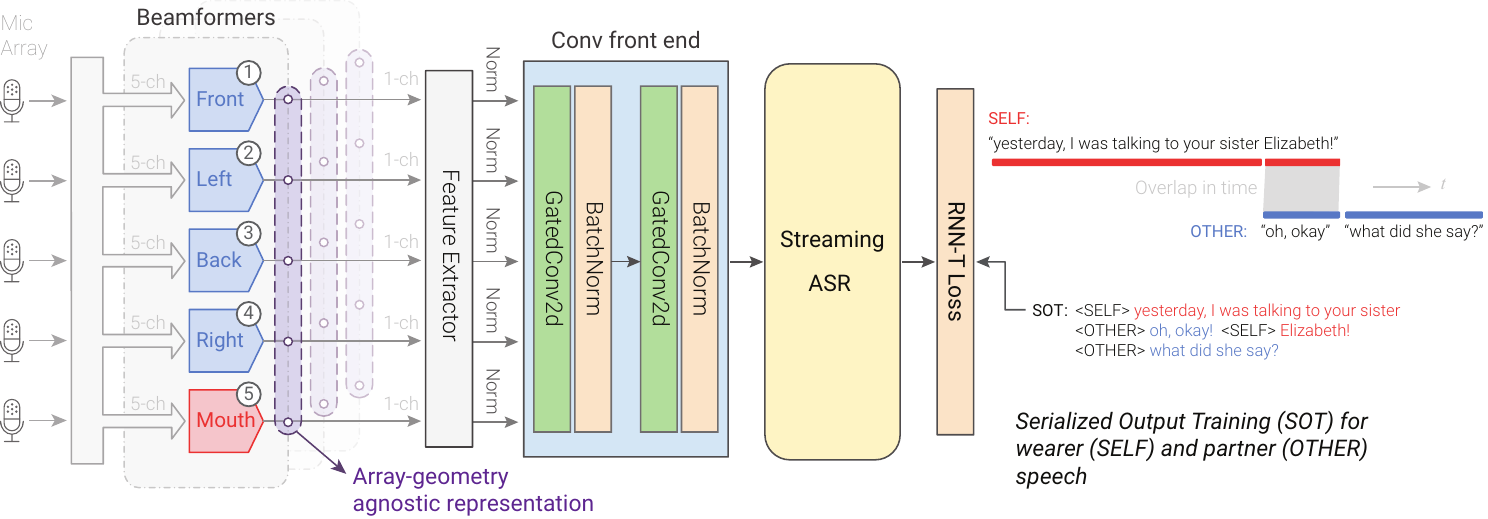}
    \vspace{-1ex}
    \caption{Proposed Array-geometry agnostic directional speech recognition architecture.}
  \label{fig:framework}
  \vspace{-3ex}
\end{figure*}

Fig.~\ref{fig:framework} illustrates the system architecture of our directional speech-recognition system. It is comprised of beamformers, feature front-end, and a streaming RNN-T based ASR system trained with serialized output training, or SOT. We will describe these components in detail in the following subsections.

\subsection{NLCMV: Non-Linearly Constrained Minimum-Variance beamforming} \label{sec:NLCMV}
Beamforming is one key component of our system for both speaker-tag detection and cross-talk suppression. Hence, our first stage is to process the raw multi-channel audio by a set of $K+1$ fixed beamformers; $K$ horizontal steering directions around the smart-glasses device plus one towards the speaker's mouth direction. These beamformers use predetermined coefficients. This converts the problem from comparing raw phase differences to one of comparing magnitudes and feature characteristics across multiple steering directions.

Our previous work~\cite{lin2023directional} used a conventional beamformer algorithm, Minimum variance distortionless response (MVDR)~\cite{capon1969high}, which aims to minimize the estimated beamformer output level while preserving the integrity of the desired signal. That approach lacks control over null directions, which can vary significantly across different frequencies, and neglects white noise during optimization. In this paper, we refine the beamformer by introducing a novel Non-Linearly Constrained Minimum Variance (NLCMV) criterion, which incorporates white noise gain and null direction control into its formulation. Specifically, NLCMV optimizes the beamformer weights $\mv{h}(jw)$ of each steering direction by minimizing
\begin{eqnarray}
    \footnotesize{
    \mv{h}^H(j\omega)
    \left[\mv{\Phi}_{dd}(j\omega) + \underbrace{\phi_{pp}(w)\sum_{n=1}^{N}\alpha_{p,n}\cdot\mv{g}_n(j\omega)\mv{g}_n^{H}(j\omega)}_{\text{soft control of null directions}}\right]\mv{h}(j\omega)
    }
\label{eq:rec}
\end{eqnarray}
which is subject to the linear equality and nonlinear inequality constraints, which are simplified to the following form:
\begin{equation}
\left\{
\begin{aligned}
  & \mv{h}^H{(j\omega)}\mv{g}{(j\omega)}=1,\\
  & c(w) \triangleq \underbrace{\mv{h}^{H}({j\omega})\mv{\Psi}(j\omega)\mv{h}(j\omega)<=0}_{\text{constraint on white noise gain.}},
\end{aligned}
\right .
\end{equation}
where $\mv{\Phi}_{dd}(jw)$ is the covariance matrix of diffuse noise, 
\[ \footnotesize{\mv{\Psi}(j\omega)\triangleq\textbf{I} - \mv{g}(j\omega)\mv{g}^{H}(j\omega) \cdot M \left/ \left[\sum_{m=1}^{M}|G_m(j\omega)|^2\right] \right. ,} \] 
The $G_m(j\omega)$ are measured channel responses from the target speech source to the $m$-th of $M$ microphones (ATFs), $N$ is the number of point noise sources, $\phi_{pp}(w)$ is the PSD of point noise, $\alpha_{p,n}$ is the $n$th point noise weight, and $\textbf{I}$ is the identity matrix.

\begin{figure}[!b]
\vspace{-4mm}
\hspace*{0.1\columnwidth}
\includegraphics[width=0.85\columnwidth]{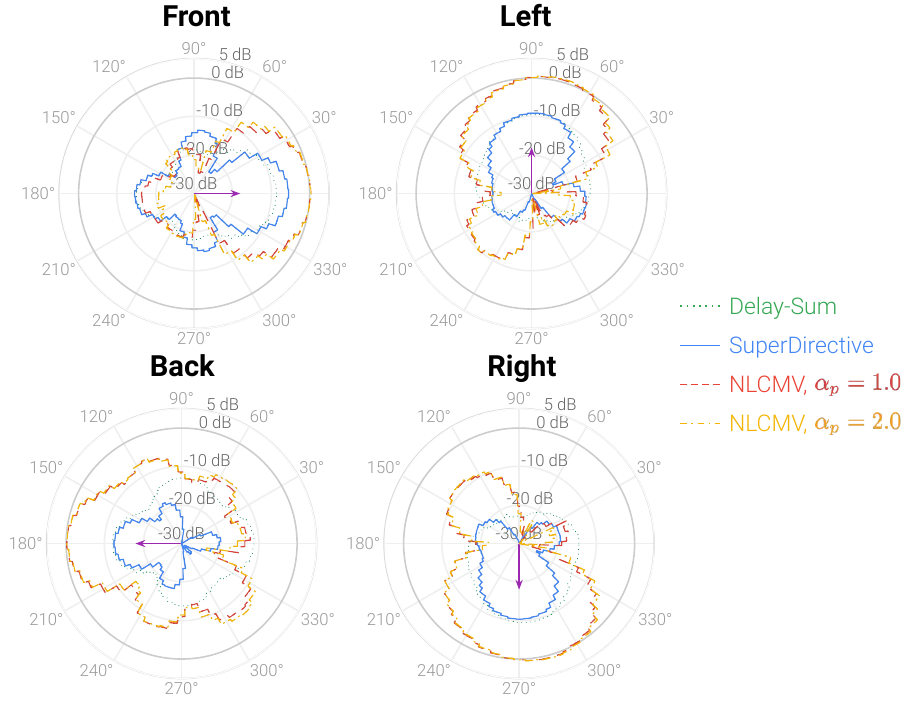}
    \vspace{-2ex}
    \caption{Beam patterns at 1000Hz for Aria glasses on 4 directions.}
  \label{fig:pattern}
\end{figure}

For illustration, Fig.~\ref{fig:pattern} compares NLCMV beam patterns to conventional delay-and-sum and super-directive ones~\cite{elko2000superdirectional,huang2016superdirective,doclo2007superdirective}. Compared to super-directive, NLCMV achieves a superior 10dB gain at the designated look direction, such as backwards, and early ASR tests on real data showed roughly a 0.7\% absolute WER gain.

\subsection{Convolutional front-end}
From the multiple channels received from the beamformers, we next extract per-channel log-Mel features (which are normalized w.r.t.~corpus mean/variance for better convergence). I.e. instead of feature vectors as in regular single-channel ASR, we have feature {\em tensors}, where the second dimension represents the steering direction. Note that log-Mel processing removes phase information which in raw audio carries the directional information. That is OK, since this information has already been perused by the beamformers, and is therefore at this point reflected as amplitude information.

Unlike our previous work~\cite{lin2023directional}, we add two convolutional blocks to further refine the extracted log-mel features. Each convolutional block is composed of a 2-D convolutional layer, succeeded by batch normalization~\cite{ioffe2015batch}, and utilizes gated linear units (GLU)~\cite{dauphin2017language} as the activation function. I.e., while our previous system~\cite{lin2023directional} just concatenated all features from all beams and linearly projected then, we now leverage a convolutional front-end aiming to retain more directional information by keeping the channels separate for a few more layers, while simultaneously reducing the feature dimension through a stride of 2. On a setup similar to this paper's results section, this improved the speaker-attributed WER by an absolute 1.3\%.

\begin{table*}[!ht]
\centering
\setlength{\tabcolsep}{2pt}
\resizebox{0.95\textwidth}{!}{
\begin{tabular}{lcccccccccccccccc}
\hline
\multicolumn{1}{c|}{\multirow{2}{*}{Model}}          & \multicolumn{1}{c|}{\multirow{2}{*}{Data}} & \multicolumn{3}{c|}{WER\%, Aria$_\mathrm{A}$}                   & \multicolumn{3}{c|}{WER\%, Comp$_\mathrm{A}$}               & \multicolumn{3}{c|}{WER\%, Comp$_\mathrm{B}$}               & \multicolumn{3}{c|}{WER\%, Comp$_\mathrm{C}$}               & \multicolumn{3}{c}{WER\%, Comp$_\mathrm{D}$} \\ \cline{3-17} 
\multicolumn{1}{c|}{}                                & \multicolumn{1}{l|}{}                      & u/a & self & \multicolumn{1}{c|}{other} & u/a & self & \multicolumn{1}{c|}{other} & u/a & self & \multicolumn{1}{c|}{other} & u/a & self & \multicolumn{1}{c|}{other} & u/a   & self   & other   \\ \hline\hline
\multicolumn{17}{c}{w/o noise and w/o bystanders}                                                                                                                                                                                                                                                                        \\ \hline
\multicolumn{1}{l|}{Matching geometry} & \multicolumn{1}{r|}{100\%}                 & 8.0     & 8.0  & \multicolumn{1}{c|}{8.1}   & 8.4     & 8.2  & \multicolumn{1}{c|}{8.6}   & 8.3     & 8.1  & \multicolumn{1}{c|}{8.4}   & 8.0     & 8.2  & \multicolumn{1}{c|}{7.9}   & 8.0       & 8.0    & 7.9     \\ \hline
\multicolumn{1}{l|}{Multi-geometry}                  & \multicolumn{1}{r|}{5$\times$20\%}                & 6.1     & 6.2  & \multicolumn{1}{c|}{6.5}   & 6.2     & 6.1  & \multicolumn{1}{c|}{6.4}   & 6.1     & 6.0  & \multicolumn{1}{c|}{6.2}   & 6.1     & 6.0  & \multicolumn{1}{c|}{6.1}   & 6.1       & 6.1    & 6.3     \\ \hline
\multicolumn{1}{l|}{Geometry-agnostic}               & \multicolumn{1}{r|}{5$\times$20\%}                & 6.3     & 6.5  & \multicolumn{1}{c|}{6.5}   & 6.3     & 6.3  & \multicolumn{1}{c|}{6.3}   & 6.0     & 6.2  & \multicolumn{1}{c|}{5.8}   & 6.1     & 6.2  & \multicolumn{1}{c|}{6.1}   & 6.2       & 6.2    & 6.2     \\ \hline\hline
\multicolumn{17}{c}{w/ noise and w/ bystanders, overlap ratio 0\%}                                                                                                                                                                                                                                                       \\ \hline
\multicolumn{1}{l|}{Matching geometry}                  & \multicolumn{1}{r|}{100\%}                 & 20.5    & 12.0 & \multicolumn{1}{c|}{27.6}  & 19.1    & 11.2 & \multicolumn{1}{c|}{25.7}  & 19.8    & 11.5 & \multicolumn{1}{c|}{26.6}  & 18.8    & 10.9 & \multicolumn{1}{c|}{25.3}  & 18.8      & 11.1   & 25.1    \\ \hline
\multicolumn{1}{l|}{Mismatching geometry}                  & \multicolumn{1}{r|}{100\%}                 &  36.5   & 53.5 & \multicolumn{1}{c|}{50.1}  &  31.1   & 18.8 & \multicolumn{1}{c|}{41.6}  &  34.4   &  22.2 & \multicolumn{1}{c|}{51.4}  &   22.0  & 12.3 & \multicolumn{1}{c|}{30.2}  &   19.6     & 11.1   &  26.6  \\ \hline
\multicolumn{1}{l|}{Multi-geometry}                  & \multicolumn{1}{r|}{5$\times$20\%}                & 16.3    & 8.9  & \multicolumn{1}{c|}{22.7}  & 16.2    & 8.4  & \multicolumn{1}{c|}{22.6}  & 15.2    & 8.1  & \multicolumn{1}{c|}{21.2}  & 15.3    & 8.2  & \multicolumn{1}{c|}{21.3}  & 15.2      & 8.2    & 21.0    \\ \hline
\multicolumn{1}{l|}{Geometry-agnostic}               & \multicolumn{1}{r|}{5$\times$20\%}                & 16.7    & 9.6  & \multicolumn{1}{c|}{23.2}  & 16.7    & 8.8  & \multicolumn{1}{c|}{22.9}  & 15.6    & 8.5  & \multicolumn{1}{c|}{21.6}  & 15.6    & 8.5  & \multicolumn{1}{c|}{21.4}  & 15.7      & 8.4    & 21.6    \\ \hline\hline
\multicolumn{17}{c}{w/ noise and w/ bystanders, overlap ratio 50\%}                                                                                                                                                                                                                                                      \\ \hline
\multicolumn{1}{l|}{Matching geometry}                  & \multicolumn{1}{r|}{100\%}                 & 21.6    & 12.6 & \multicolumn{1}{c|}{28.9}  & 20.5    & 11.7 & \multicolumn{1}{c|}{27.9}  & 21.2    & 12.1 & \multicolumn{1}{c|}{28.6}  & 19.6    & 11.2 & \multicolumn{1}{c|}{26.5}  & 20.5      & 11.7   & 27.9    \\ \hline
\multicolumn{1}{l|}{Multi-geometry}                  & \multicolumn{1}{r|}{5$\times$20\%}                & 17.0    & 9.4  & \multicolumn{1}{c|}{23.5}  & 17.3    & 8.7  & \multicolumn{1}{c|}{24.4}  & 16.4    & 8.5  & \multicolumn{1}{c|}{23.0}  & 16.0    & 8.4  & \multicolumn{1}{c|}{22.3}  & 16.3      & 8.4    & 22.7    \\ \hline
\multicolumn{1}{l|}{Geometry-agnostic}               & \multicolumn{1}{r|}{5$\times$20\%}                & 17.6    & 9.9  & \multicolumn{1}{c|}{24.2}  & 17.8    & 9.1  & \multicolumn{1}{c|}{24.9}  & 16.8    & 8.9  & \multicolumn{1}{c|}{23.4}  & 16.2    & 8.6  & \multicolumn{1}{c|}{22.6}  & 16.6      & 8.6    & 23.1    \\ \hline
\end{tabular}
}
\caption{Speaker un-attributed ("u/a") and attributed ("self", "other") word error rates (WER) on simulated test data for five different array geometries, with "Matching geometry" (same array in training and test), "Multi-geometry" (multiple geometries with matching array-id embedding, applied to 20\% of the data, resp.), and "Geometry-agnostic" (multiple without array id). "Mismatching geometry" uses a model trained on the respective geometry one column to the left (or on Comp$_\mathrm{D}$ for Aria$_\mathrm{A}$).
}
\vspace{-5mm}
\label{table:simu}
\end{table*}

\subsection{Streaming ASR with Serialized Output Training}
Our streaming ASR model is the same as~\cite{lin2023directional}: a Neural Transducer~\cite{mahadeokar2021alignment,moritz2022investigation,sainath2020streaming,li2020developing}, specifically a Recurrent Neural Network Transducer, or RNN-T, that consists of three components: an encoder, a prediction network, and a joiner network. There is no external language model. As in~\cite{lin2023directional}, multi-talker overlapped speech is handled via {\em serialized output training}, or SOT~\cite{kanda2022streaming,Chang2022extGTC}, where the model is trained to insert tags marking speaker changes---in our case between the wearer and a target speaker (other). The training process uses the "alignment-restricted RNN-T" (AR-RNN-T) technique~\cite{mahadeokar2021alignment} for acceleration.

\section{Experiments and Results}

\subsection{Dataset}

Models are trained on an in-house dataset of 14.6k hours of de-identified video data that is publicly shared by Facebook users---single-channel audio. As real multi-channel training data of sufficient amounts is not available, all multi-channel training data for all microphone-array geometries must be simulated. We first generate 1M multi-channel room impulse responses (RIRs) using image-source methods (ISM)\cite{lehmann2008prediction} via the ``pyroomacoustics'' library~\cite{scheibler2018pyroomacoustics}. Room sizes range from [5, 5, 2] to [10, 10, 6] meters. We then simulate training data by placing single-channel audio clips in space as the wearer ("self"), the conversation partner ("other"), and unrelated bystanders, simulating a conversation between self and other with some overlap, and bystander crosstalk. The "other" speech is located at forward-facing angles of -60 to +60$^{\circ}$, while the bystander is positioned at random locations outside that range (i.e.~left, right, or behind the wearer). (In~\cite{lin2023directional}, this configuration is labeled V4.)

We evaluate our proposed methods on both real and simulated test sets. The simulated set consists of an additional 3.7 hours of in-house video, converted to multi-channel via simulation like the training data, except using different simulated RIRs. Additionally, real test data was collected consisting of conversations between a wearer wearing Project Aria prototyping glasses (Section~\ref{sec:devices}) and a conversation partner at a distance of around 4 to 6 feet. All data is bilingual ("self" speaks English while "other" speaks Spanish).  

Lastly, noise from the DNS Challenge~\cite{reddy2020interspeech} was added to the clean audio segments in training and test, at SNRs ranging from $-5$ to 30 dB w.r.t.~the combined audio of wearer and partner, at intervals of 1 dB. Three overlap configurations between bystanders and main speakers are investigated: no crosstalk, crosstalk not overlapping (0\%), and 50\% overlap with the main speakers (self or other).
\subsection{Devices} \label{sec:devices}

Two hardware devices were used in this work, the publicly available Project Aria glasses~\cite{somasundaram2023project} and a composite hardware prototype that combines several microphone geometries for evaluating microphone placements. For both, measurements of Acoustic Transfer Functions (ATFs) for all microphones were available to us and were used for the beamformer design (Section \ref{sec:NLCMV}). Unfortunately, unlike Aria, the composite prototype is mechanically not suitable for collecting real conversations, relegating us to simulated test data for it.

\begin{figure}[!b]
\vspace{-2mm}
    \centering
    \includegraphics[width=0.62\linewidth]{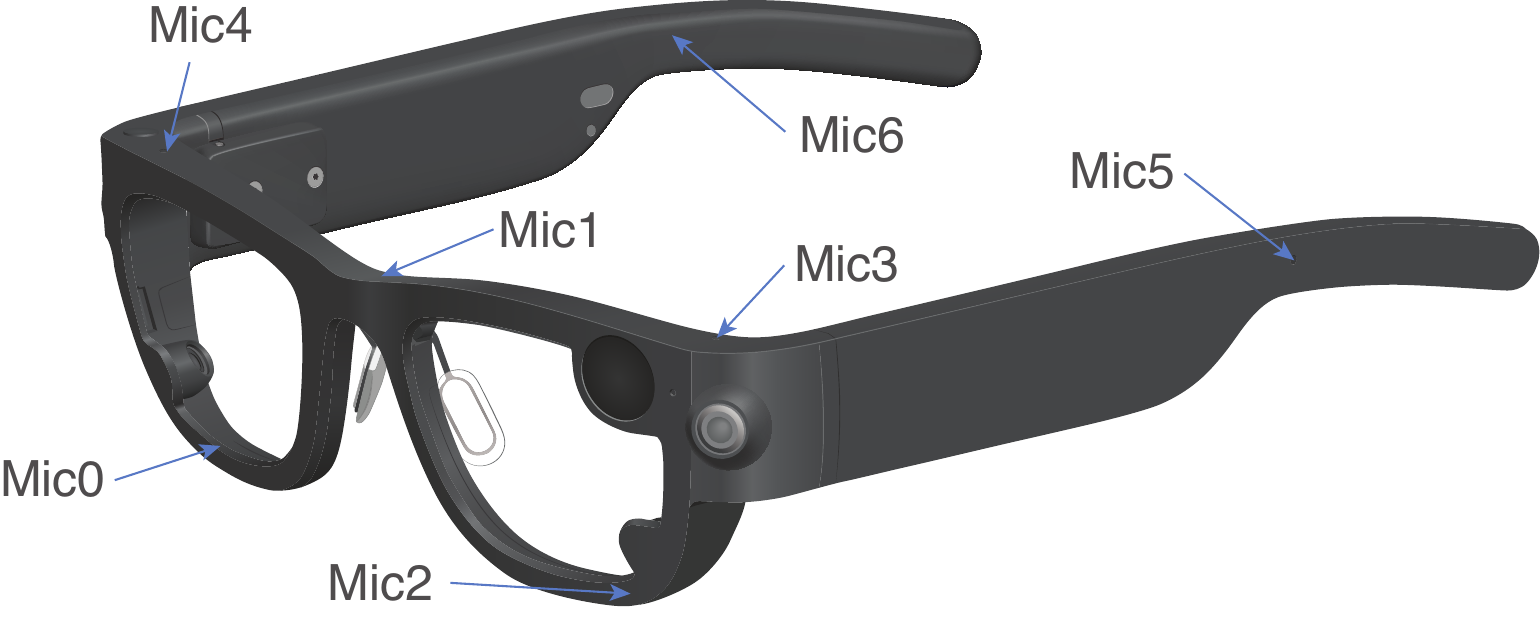}
    \caption{Microphone locations on Project Aria glasses~\cite{somasundaram2023project}.}
    \label{fig:glasses}
\end{figure}

For our application, we target microphone arrays of 5 channels. Both hardware devices have more microphones than that. This way, we can experiment with multiple 5-channel configurations by dropping different sets of microphones. We define two 5-channel subsets for Aria named {Aria$_\mathrm{A}$} (seen in training, using Mic2, Mic3, Mic4, Mic5, and Mic6 per Fig.~\ref{fig:glasses}) and {Aria$_\mathrm{B}$} (not seen in training, substituting nose Mic0 in place of Mic2). From the composite prototype, we derive five meaningful configurations labeled Comp$_\mathrm{A}$, Comp$_\mathrm{B}$, Comp$_\mathrm{C}$, and Comp$_\mathrm{D}$ (seen in training) as well as and Comp$_\mathrm{E}$ (not seen in training), which differ to the order of several cm in where on the temple arm microphones are placed, as well as nose-microphone location.
The 4-channel configuration in the contrast experiment in Section \ref{sec:4mic} is based on configuration A except that the nose microphone is dropped entirely, leaving only 4 channels.

\subsection{Model configuration}

The model configuration is similar to~\cite{lin2023directional}. For each beamformer direction, 80-dimensional log-Mel filterbank features are extracted. Input features from all channels (steering directions) are then fed into the Convolutional front-end, which consists of 2 conv2d blocks each with 5 channels, filters of size \sxs{2}{5} and a stride setting of \sxs{1}{2}. Then, six consecutive frames are stacked to form a 320-dimensional vector, reducing the sequence length by 6x. This is followed by 20 Emformer layers~\cite{shi2021emformer}, each with 4 attention heads and 2048-dimensional feed-forward layers. The RNN-T's {\em prediction network} contains one 256-dimensional LSTM layer with layer normalization and dropout. Lastly, the encoder and predictor outputs are both projected to 768 dimensions and passed to an additive {\em joiner network}, which contains a ReLU followed by linear layer with 9001 output SentencePiece-based units.

Furthermore, for the "multi-geometry" system trained on multiple geometries, we incorporate array ids encoded as a one-hot embedding that gets concatenated with the output of the convolutional front-end. The array-id is used to switch beamformer parameters. Such system can distinguish multiple devices used during training, but does not support previously unseen devices. On the other hand, the "Geometry-agnostic" variant is trained on the same multiple geometries but without array ids, remaining adaptable for handling previously unseen devices. We want to clarify that "agnostic" is in terms of the ASR model, not the beamformers which are still created for the actual target device, seen in training or not.

Lastly, all models are trained for 8 epochs, with an Adam$_\mathrm{sam}$ optimizer, a tri-stage learning-rate scheduler with a base learning rate of 0.0005, and a warmup of 10,000 batches.
%
\begin{table}[]
\resizebox{\columnwidth}{!}{
\setlength{\tabcolsep}{2pt}
\begin{tabular}{l|c|c|ccc}
\hline
\multirow{2}{*}{Model Type} & \multirow{2}{*}{Data} & \multirow{2}{*}{Test Device} & \multicolumn{3}{c}{WER\%} \\ \cline{4-6} 
                            &                       &                              & u/a  & self  & other \\ \hline\hline
Matching geometry              & \multicolumn{1}{r|}{100\%}  & Aria$_\mathrm{A}$                       & 22.9     & 13.3  & 26.1  \\ \hline
Mismatching geometry              & \multicolumn{1}{r|}{100\%}  & Aria$_\mathrm{B}$                       & 23.0     & 16.4  & 27.7  \\ \hline
Multi-geometry              & 5$\times$20\%                & Aria$_\mathrm{A}$                        & 20.1     & 10.0  & 21.8  \\ \hline
Geometry-agnostic           & 5$\times$20\%                & Aria$_\mathrm{A}$                        & 20.4     & 10.1  & 22.2  \\ \hline
\end{tabular}
}
\caption{Word error rates on the real test dataset.}
\label{table:real}
\vspace{-3mm}
\end{table}
\begin{table}[!b]
\centering
\setlength{\tabcolsep}{3pt}
\begin{tabular}{l|c|c|ccc}
\hline
\multirow{2}{*}{Test Device} &Seen/& \multirow{2}{*}{Data Type} & \multicolumn{3}{c}{WER\%} \\ \cline{4-6}
                             &Unseen&                            & u/a  & self  & other \\ \hline\hline
Aria$_\mathrm{A}$ & seen                 & real                       & 20.4     & 10.1  & 22.2 
\\ \hline
Aria$_\mathrm{B}$ & unseen         & real                       & 20.7     & 10.1  & 22.8 \\ \hline\hline
Comp$_\mathrm{B}$ & seen          & simulated                 & 15.6     & 8.5   & 21.6  \\ \hline
Comp$_\mathrm{D}$ & seen          & simulated                 & 15.7     & 8.4   & 21.6  \\ \hline
Comp$_\mathrm{E}$ & unseen           & simulated                       & 15.9     & 8.5   & 22.0  \\ \hline\hline
Comp$_\mathrm{A}$ & seen           & simulated              & 16.7     & 8.8  & 22.9  \\ \hline
Comp$_\mathrm{A,4mic}$ & unseen    & simulated                        & 26.0     & 27.9  & 32.6  \\ \hline
\end{tabular}
\caption{Performance in terms of WER on seen vs.~unseen devices, for the "Geometry-agnostic" model which did not include "Unseen" device geometries in the training. Noise and bystanders are added for the simulated test sets and overlap ratio is 0\%.}
\label{table:unseen}
\end{table}

\subsection{Results}

All results show two types of WER: speaker-unattributed (denoted "u/a") and speaker-attributed (denoted "self" and "other"). "u/a" scores the sequence of words, irrespective of which speaker they were attributed to, while "self" and "other" score only words attributed to the respective speaker in ASR output and reference. The "u/a" metric is {\em not} the average of "self" and "other"---a word attributed to the wrong speaker counts as an insertion for one speaker and a deletion for the other.

\subsubsection{Training on multiple geometries, test devices seen in training}
Table~\ref{table:simu} shows results on simulated test data, which we can create for all relevant combinations. First, we see that training on multiple geometries at once ("Multi-geometry" and "Geometry-agnostic") not only works (the original purpose of this work), but outperforms training on matched geometries only, by as much as 28\% relative (e.g.~from 8.3\% to 6.0\% for the clean Comp$_\mathrm{B}$/"Geometry-agnostic"). We speculate that the incorporation of more devices/geometries in the data simulation contributes to the robustness, e.g.~discouraging the model from over-indexing to fine structure in the beam patterns.

Secondly, compared to "Multi-geometry," the exclusion of array-id information, with the goal of being "Geometry-agnostic" model, led to only a slight WER increases bounded by roughly 0.5\% absolute with few exceptions. This is consistent across three different settings, e.g.~with and without bystanders.

Similar results are shown in Table \ref{table:real}, but for real data instead. The method generalizes well to real data, achieving a 2.5\% absolute gain by going from matching geometry to "Geometry-agnostic."
\vspace{-2mm}
\subsubsection{Geometry-agnostic model with unseen devices}
How about unseen geometries? In Table~\ref{table:simu}, shows under "Mismatched geometry" a drastic accuracy hit for models trained on one geometry but naively tested on another, with WERs of almost 40\%.

This is, however, not so if we train on multiple geometries. Table~\ref{table:unseen} shows WERs for the "Geometry-agnostic" model when tested with devices not seen vs.~seen in training. In the first two sections (Aria$_\mathrm{A}$ (seen) vs.~Aria$_\mathrm{B}$ (unseen) real data; Comp$_\mathrm{B}$/Comp$_\mathrm{D}$ (seen) vs. Comp$_\mathrm{E}$ (unseen) simulated data), WERs deviate by no more than 0.6\% absolute. (Both Aria$_\mathrm{A}$ vs.~Aria$_\mathrm{B}$ and Comp$_\mathrm{B}$ vs.~Comp$_\mathrm{E}$ differ only in the nose microphone, while Comp$_\mathrm{D}$ and Comp$_\mathrm{E}$ differ in three microphones, but note that moving even one microphone changes all beamformer weights.) 

In this condition, {\em the model is indeed geometry-agnostic}. Although not yet tested for explicitly, this also gives some confidence that the Geometry-agnostic system will robustly accommodate variations in head sizes/shapes, hair, headwear, etc.

\label{sec:4mic}
We also tested a more extreme case, simulating the situation where system designers decide to drop the nose microphone altogether, denoted by "Comp$_\mathrm{A}$ (4-mic)". Here, the method reaches its limits: This significant deviation from the 5-channel geometries used during training causes a noticeable drop in performance, pushing all WERs above 25\%. The goal of agnosticity is not achieved here. Maybe one should not expect this to work in the first place, as there is nothing in beamformer objective to explicitly encourage beamformers across geometries to be similar. Investigating such a constraint is future work.

\section{Conclusion}
This paper addresses an important practical problem of microphone arrays being a "moving target" during system development. We propose a first step towards Array-Geometry Agnostic Directional Speech Recognition (AGADIR): As long as geometry variations are moving around microphones by a few mm to cm and do not change the fundamental nature of the array, we find that training the directional ASR model with multiple geometries not only works but also generalizes to new unseen variations, indeed exhibiting the desired geometry-agnostic behavior in this case. Furthermore, it improves the baseline WER by on the order of 20\% relative (up to 28\%). However, more work is needed to achieve agnosticity to more extreme geometry variations such as dropping a microphone altogether, possibly via an additional constraint to explicitly keep beamformers consistent across geometries. In addition, the paper introduces an innovative beamformer design tailored for directional speech recognition, demonstrating superiority over conventional methods.

{\footnotesize
\bibliographystyle{IEEEbib}
\bibliography{strings,refs}}

\begin{thebibliography}{10}

\bibitem{lin2023directional}
Ju~Lin, Niko Moritz, Ruiming Xie, Kaustubh Kalgaonkar, Christian Fuegen, and
  Frank Seide,
\newblock ``Directional speech recognition for speaker disambiguation and
  cross-talk suppression,''
\newblock {\em Proc. INTERSPEECH 2023}, pp. 3522--3526, 2023.

\bibitem{taherian2022one}
Hassan Taherian, Sefik~Emre Eskimez, Takuya Yoshioka, Huaming Wang, Zhuo Chen,
  and Xuedong Huang,
\newblock ``One model to enhance them all: array geometry agnostic
  multi-channel personalized speech enhancement,''
\newblock in {\em ICASSP 2022-2022 IEEE International Conference on Acoustics,
  Speech and Signal Processing (ICASSP)}. IEEE, 2022, pp. 271--275.

\bibitem{yoshioka2022vararray}
Takuya Yoshioka, Xiaofei Wang, Dongmei Wang, Min Tang, Zirun Zhu, Zhuo Chen,
  and Naoyuki Kanda,
\newblock ``Vararray: Array-geometry-agnostic continuous speech separation,''
\newblock in {\em ICASSP 2022-2022 IEEE International Conference on Acoustics,
  Speech and Signal Processing (ICASSP)}. IEEE, 2022, pp. 6027--6031.

\bibitem{kanda2023vararray}
Naoyuki Kanda, Jian Wu, Xiaofei Wang, Zhuo Chen, Jinyu Li, and Takuya Yoshioka,
\newblock ``Vararray meets t-sot: Advancing the state of the art of streaming
  distant conversational speech recognition,''
\newblock in {\em ICASSP 2023-2023 IEEE International Conference on Acoustics,
  Speech and Signal Processing (ICASSP)}. IEEE, 2023, pp. 1--5.

\bibitem{chang2019mimo}
Xuankai Chang, Wangyou Zhang, Yanmin Qian, Jonathan Le~Roux, and Shinji
  Watanabe,
\newblock ``{MIMO}-speech: {E}nd-to-end multi-channel multi-speaker speech
  recognition,''
\newblock in {\em 2019 IEEE Automatic Speech Recognition and Understanding
  Workshop (ASRU)}. IEEE, 2019, pp. 237--244.

\bibitem{wang2018multi}
Zhong-Qiu Wang, Jonathan Le~Roux, and John~R Hershey,
\newblock ``Multi-channel deep clustering: {D}iscriminative spectral and
  spatial embeddings for speaker-independent speech separation,''
\newblock in {\em 2018 IEEE International Conference on Acoustics, Speech and
  Signal Processing (ICASSP)}. IEEE, 2018, pp. 1--5.

\bibitem{shao2022multi}
Yiwen Shao, Shi-Xiong Zhang, and Dong Yu,
\newblock ``Multi-channel multi-speaker {ASR} using {3D} spatial feature,''
\newblock in {\em ICASSP 2022-2022 IEEE International Conference on Acoustics,
  Speech and Signal Processing (ICASSP)}. IEEE, 2022, pp. 6067--6071.

\bibitem{chen2018multi}
Zhuo Chen, Xiong Xiao, Takuya Yoshioka, Hakan Erdogan, Jinyu Li, and Yifan
  Gong,
\newblock ``Multi-channel overlapped speech recognition with location guided
  speech extraction network,''
\newblock in {\em 2018 IEEE Spoken Language Technology Workshop (SLT)}. IEEE,
  2018, pp. 558--565.

\bibitem{sainath2016factored}
Tara~N Sainath, Ron~J Weiss, Kevin~W Wilson, Arun Narayanan, and Michiel
  Bacchiani,
\newblock ``Factored spatial and spectral multichannel raw waveform cldnns,''
\newblock in {\em 2016 IEEE International Conference on Acoustics, Speech and
  Signal Processing (ICASSP)}. IEEE, 2016, pp. 5075--5079.

\bibitem{he2020spatial}
Weipeng He, Lu~Lu, Biqiao Zhang, Jay Mahadeokar, Kaustubh Kalgaonkar, and
  Christian Fuegen,
\newblock ``Spatial attention for far-field speech recognition with deep
  beamforming neural networks,''
\newblock in {\em ICASSP 2020-2020 IEEE International Conference on Acoustics,
  Speech and Signal Processing (ICASSP)}. IEEE, 2020, pp. 7499--7503.

\bibitem{capon1969high}
Jack Capon,
\newblock ``High-resolution frequency-wavenumber spectrum analysis,''
\newblock {\em Proceedings of the IEEE}, vol. 57, no. 8, pp. 1408--1418, 1969.

\bibitem{elko2000superdirectional}
Gary~W Elko, SL~Gay, and J~Benesty,
\newblock ``Superdirectional microphone arrays,''
\newblock {\em Kluwer International Series in Engineering and Computer
  Science}, pp. 181--238, 2000.

\bibitem{huang2016superdirective}
Gongping Huang, Jacob Benesty, and Jingdong Chen,
\newblock ``Superdirective beamforming based on the krylov matrix,''
\newblock {\em IEEE/ACM Transactions on Audio, Speech, and Language
  Processing}, vol. 24, no. 12, pp. 2531--2543, 2016.

\bibitem{doclo2007superdirective}
Simon Doclo and Marc Moonen,
\newblock ``Superdirective beamforming robust against microphone mismatch,''
\newblock {\em IEEE Transactions on Audio, Speech, and Language Processing},
  vol. 15, no. 2, pp. 617--631, 2007.

\bibitem{ioffe2015batch}
Sergey Ioffe and Christian Szegedy,
\newblock ``Batch normalization: Accelerating deep network training by reducing
  internal covariate shift,''
\newblock in {\em International conference on machine learning}. PMLR, 2015,
  pp. 448--456.

\bibitem{dauphin2017language}
Yann~N Dauphin, Angela Fan, Michael Auli, and David Grangier,
\newblock ``Language modeling with gated convolutional networks,''
\newblock in {\em International conference on machine learning}. PMLR, 2017,
  pp. 933--941.

\bibitem{mahadeokar2021alignment}
Jay Mahadeokar, Yuan Shangguan, Duc Le, Gil Keren, Hang Su, Thong Le,
  Ching-Feng Yeh, Christian Fuegen, and Michael~L Seltzer,
\newblock ``Alignment restricted streaming recurrent neural network
  transducer,''
\newblock in {\em 2021 IEEE Spoken Language Technology Workshop (SLT)}. IEEE,
  2021, pp. 52--59.

\bibitem{moritz2022investigation}
Niko Moritz, Frank Seide, Duc Le, Jay Mahadeokar, and Christian Fuegen,
\newblock ``An investigation of monotonic transducers for large-scale automatic
  speech recognition,''
\newblock {\em arXiv preprint arXiv:2204.08858}, 2022.

\bibitem{sainath2020streaming}
Tara~N Sainath, Yanzhang He, Bo~Li, Arun Narayanan, Ruoming Pang, Antoine
  Bruguier, Shuo-yiin Chang, Wei Li, Raziel Alvarez, Zhifeng Chen, et~al.,
\newblock ``A streaming on-device end-to-end model surpassing server-side
  conventional model quality and latency,''
\newblock in {\em ICASSP 2020-2020 IEEE International Conference on Acoustics,
  Speech and Signal Processing (ICASSP)}. IEEE, 2020, pp. 6059--6063.

\bibitem{li2020developing}
Jinyu Li, Rui Zhao, Zhong Meng, Yanqing Liu, Wenning Wei, Sarangarajan
  Parthasarathy, Vadim Mazalov, Zhenghao Wang, Lei He, Sheng Zhao, et~al.,
\newblock ``Developing rnn-t models surpassing high-performance hybrid models
  with customization capability,''
\newblock {\em arXiv preprint arXiv:2007.15188}, 2020.

\bibitem{kanda2022streaming}
Naoyuki Kanda, Jian Wu, Yu~Wu, Xiong Xiao, Zhong Meng, Xiaofei Wang, Yashesh
  Gaur, Zhuo Chen, Jinyu Li, and Takuya Yoshioka,
\newblock ``Streaming multi-talker {ASR} with token-level serialized output
  training,''
\newblock {\em arXiv preprint arXiv:2202.00842}, 2022.

\bibitem{Chang2022extGTC}
Xuankai Chang, Niko Moritz, Takaaki Hori, Shinji Watanabe, and Jonathan~Le
  Roux,
\newblock ``Extended graph temporal classification for multi-speaker end-to-end
  {ASR},''
\newblock in {\em 2022 IEEE International Conference on Acoustics, Speech and
  Signal Processing (ICASSP)}, 2022, pp. 7322--7326.

\bibitem{lehmann2008prediction}
Eric~A Lehmann and Anders~M Johansson,
\newblock ``Prediction of energy decay in room impulse responses simulated with
  an image-source model,''
\newblock {\em The Journal of the Acoustical Society of America}, vol. 124, no.
  1, pp. 269--277, 2008.

\bibitem{scheibler2018pyroomacoustics}
Robin Scheibler, Eric Bezzam, and Ivan Dokmani{\'c},
\newblock ``Pyroomacoustics: A python package for audio room simulation and
  array processing algorithms,''
\newblock in {\em 2018 IEEE international conference on acoustics, speech and
  signal processing (ICASSP)}. IEEE, 2018, pp. 351--355.

\bibitem{reddy2020interspeech}
Chandan~KA Reddy, Vishak Gopal, Ross Cutler, Ebrahim Beyrami, Roger Cheng,
  Harishchandra Dubey, Sergiy Matusevych, Robert Aichner, Ashkan Aazami,
  Sebastian Braun, et~al.,
\newblock ``The interspeech 2020 deep noise suppression challenge: Datasets,
  subjective testing framework, and challenge results,''
\newblock in {\em INTERSPEECH}, 2020.

\bibitem{somasundaram2023project}
Kiran Somasundaram, Jing Dong, Huixuan Tang, Julian Straub, Mingfei Yan,
  Michael Goesele, et~al.,
\newblock ``Project aria: A new tool for egocentric multi-modal ai research,''
\newblock {\em arXiv preprint arXiv:2308.13561}, 2023.

\bibitem{shi2021emformer}
Yangyang Shi, Yongqiang Wang, Chunyang Wu, Ching-Feng Yeh, Julian Chan, Frank
  Zhang, Duc Le, and Mike Seltzer,
\newblock ``Emformer: {E}fficient memory transformer based acoustic model for
  low latency streaming speech recognition,''
\newblock in {\em ICASSP 2021-2021 IEEE International Conference on Acoustics,
  Speech and Signal Processing (ICASSP)}. IEEE, 2021, pp. 6783--6787.

\end{thebibliography}

\end{document}